# A Local Measurement Based Comprehensive Protection Scheme for AC Microgrid

Sindhura Gupta
*Department of Electrical Engineering*
*Netaji Subhash Engineering College*
*Kolkata, India*
sindhura.puja@gmail.com

Susovan Mukhopadhyay
*Department of Electrical Engineering*
*Netaji Subhash Engineering College*
*Kolkata, India*
susovanmuk@gmail.com

Ambarnath Banerji
*Department of Electrical Engineering*
*Narula Institute of Technology*
*Kolkata, India*
ambarnathbanerji@gmail.com

Sujit K. Biswas
*Department of Electrical Engineering*
*St. Thomas' College of Engineering and Technology*
*Kolkata, India*
sujit_biswas@hotmail.com

Prasun Sanki
*Department of Electrical Engineering*
*Netaji Subhash Engineering College*
*Kolkata, India*
prasunsanki@gmail.com

*Abstract*- The popularity of low-voltage ac distribution networks is increasing day by day. However, an efficient protection scheme for low-voltage ac distribution systems is still challenging. This paper introduces a protection scheme suitable for low-voltage grid connected and islanded ac microgrid based on local measurements in order to locate, identify and isolate faults. Here current decomposition method is specifically incorporated to accomplish fault type identification. MATLAB/SIMULINK platform is chosen to examine the performance of the proposed scheme both in grid connected and islanded low-voltage ac microgrid. The feasibility of the protection scheme is extensively investigated by simulating all types of faults with substantial variations like different fault location, different fault resistances etc. The test results ensure that the proposed protection scheme is sufficiently reliable and faster for providing complete protection in low-voltage ac microgrid.

*Keywords*- low-voltage ac microgrid, fault detection, fault identification, fault isolation.

## I. INTRODUCTION

In the last decades, the concept of distributed energy resources (DERs) has become best alternative of conventional generators in order to satisfy the ever increasing electricity demand with improved reliability and much reduced distribution power losses. Additionally, DER integration in microgrid (MG) provides ultimate solution to the concern of environmental degradation [1]. Despite of several benefits, DER integrated MG suffers from the issue of protection. The available protection measures, like over-current, differential, distance-based protection etc. is not sufficient for DER integrated MG [2]. Thus concrete relaying scheme is required to ensure overall protection in MG under all operating conditions.

Generally, the level of fault current magnitude in grid connected MG is comparatively higher than islanded MG operation. Since grid consists of significantly high short circuit megavolt ampere (MVA) rating compared to local DERs, the fault current contribution from grid is higher. It has been observed that inverter based DERs can contribute maximum twice to the rated current of inverter to the fault. On the other hand synchronous or doubly fed induction generator based DERs can contribute 4 to 10 times more than inverter based DER to fault [3]. Thus under such diversified operating condition, a variation of the fault current is experienced by a relay when fault occurs at a specific location in the MG. Traditional over-current relays with fixed setting are inadequate for sensitive and selective operation of MG operating both in grid connected and islanded mode operation [4]. Thus till date many authors have proposed advanced over-current relay based protection scheme for DER integrated MG. Like over-current relay with auto updated tripping characteristics is proposed in [5]. Over-current relay with voltage restriction in [6] and adaptive over-current relay in [7] is incorporated for protection. Recently an updated over current protection is proposed in [8]. Besides, differential protection schemes like feature extraction based differential protection scheme in [9], energy based differential protection scheme in [10] are also proposed in different MG scenarios. The main concerning issue of these protection schemes is either complexity or dependency of communication network or large computational burden.

It is being observed that the above discussed references have not designed a comprehensive protection scheme where all possible MG operating conditions are considered. In addition to that majority of these topologies are either applicable for either grid connected or islanded MG operation. Impact of numerous critical no fault conditions are considered for a few articles. Based on the critical research gap the major contribution of the article is furnished below

1. The paper proposes a simple and fast comprehensive protection scheme which is capable to locate, isolate and identify the exact fault type in grid connected and islanded MG condition.
2. Here local measurement of voltage and current are mainly involved for locating the fault. A pre-defined threshold value of voltage and current is considered to identify the fault location. The threshold of voltage and current is determined after examining various critical operating conditions.

3. The proposed scheme is examined on a 415V grid connected mesh configured MG on MATLAB/SIMULINK.
4. Fast response time for fault isolation is achieved as solid state relay are incorporated instead of over-current or differential relay. Similarly static switches are used in place of circuit breakers.

The organization of the paper is as follows. The detailed fault current calculation is presented in Scetion-II. Section-III represents the fault identification and isolation procedure in detail. The overview of low-voltage ac MG is included in Section-IV. The simulated test result with detailed discussion is presented in Section-V. Finally concluding remarks are drawn in Section-VI.

## II. FAULT CURRENT CALCULATION

This article introduces a new fault identification and selective isolation topology based on local measurements of voltage and current. The proposed scheme is equally capable to identify both symmetrical and unsymmetrical faults. The detailed fault identification topology based on local voltage and current measurement is discussed in the following section.

### a) Internal fault

Here the fault is considered to be occurring inside the DER integrated ac MG. In order to analyze the internal fault a small ac MG system is considered as depicted in Fig.1. The system is considered to be consisting of two inter connected buses. The analysis of unsymmetrical and symmetrical fault condition is carried out separately as the behavior of fault current and voltage are different for the two cases.

### 1. Unsymmetrical fault condition

Decomposition of system current and voltages into their respective symmetrical components becomes essential [11] during unsymmetrical fault condition. The corresponding decomposition of system three phase current $I_a, I_b, I_c$ and three phase voltage $V_a, V_b, V_c$ under unsymmetrical fault condition are (1) and (2).

$$\begin{bmatrix} I^+ \\ I^- \\ I^0 \end{bmatrix} = \frac{1}{3} \begin{bmatrix} 1 & \alpha & \alpha^2 \\ 1 & \alpha^2 & \alpha \\ 1 & 1 & 1 \end{bmatrix} \begin{bmatrix} I_a \\ I_b \\ I_c \end{bmatrix} (1) \quad \begin{bmatrix} V^+ \\ V^- \\ V^0 \end{bmatrix} = \frac{1}{3} \begin{bmatrix} 1 & \alpha & \alpha^2 \\ 1 & \alpha^2 & \alpha \\ 1 & 1 & 1 \end{bmatrix} \begin{bmatrix} V_a \\ V_b \\ V_c \end{bmatrix} (2)$$

Here, $I^+, I^-, I^0, V^+, V^-, V^0$ represents the positive, negative and zero sequence components of the three phase current and voltages respectively. $\alpha$ is known as operator having value of $1 \angle 120°$ Following the same According to Fig.1 the positive, negative, zero sequence components of current and voltages at the buses are represented as $I_1^+, I_1^-, I_1^0$, $I_2^+, I_2^-, I_2^0$ and $V_1^+, V_1^-, V_1^0$, $V_2^+, V_2^-, V_2^0$ respectively. At the exact fault location the positive, negative, zero sequence components of the fault current and voltage are $I_f^+, I_f^-, I_f^0$ and $V_{fp}^+, V_{fp}^-, V_{fp}^0$. $Z_{s1}^+, Z_{s1}^-, Z_{s1}^0$ and $Z_{s2}^+, Z_{s2}^-, Z_{s2}^0$ are the respective sequence components of the source impedances $Z_{s1}$ and $Z_{s2}$. Likewise, the distribution line impedance sequence components are $Z_{d1}^+, Z_{d1}^-, Z_{d1}^0$ and $Z_{d2}^+, Z_{d2}^-, Z_{d2}^0$. $Z_f$ is considered as the fault impedance. According to Fig.1 the total line impedance of the three sequence network is (3)

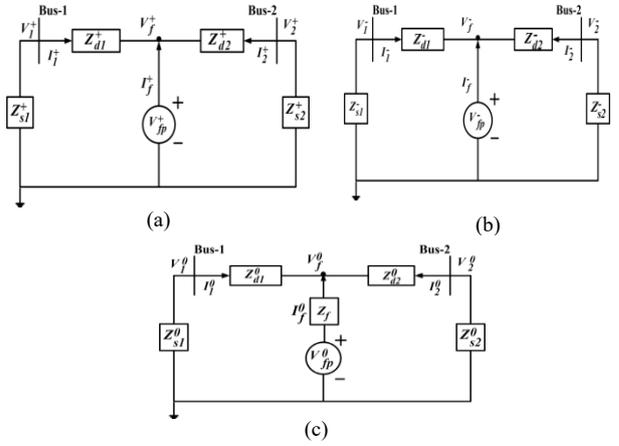

Fig.1 (a) Positive sequence network for internal fault condition, (b) negative sequence network for internal fault condition and (c) zero sequence network for internal fault condition.

$$\left.\begin{array}{l} Z_d^+ = Z_{d1}^+ + Z_{d2}^+ \\ Z_d^- = Z_{d1}^- + Z_{d2}^- \\ Z_d^0 = Z_{d1}^0 + Z_{d2}^0 \end{array}\right\} (3)$$

$$\left.\begin{array}{l} I_f^+ = -\dfrac{V_{fp}^+}{(Z_{s1}^+ + Z_{d1}^+) \| (Z_{s2}^+ + Z_{d2}^+)} \\ I_f^- = -\dfrac{V_{fp}^-}{(Z_{s1}^- + Z_{d1}^-) \| (Z_{s2}^- + Z_{d2}^-)} \\ I_f^0 = -\dfrac{V_{fp}^0}{3Z_f + (Z_{s1}^0 + Z_{d1}^0) \| (Z_{s2}^0 + Z_{d2}^0)} \end{array}\right\} (4)$$

According to the decomposition process the formulated sequence components of the fault current are presented as (4). Assuming, $Z_1^+ = Z_{s1}^+ + Z_{d1}^+$, $Z_1^- = Z_{s1}^- + Z_{d1}^-$ $Z_1^0 = Z_{s1}^0 + Z_{d1}^0$, $Z_2^+ = Z_{s2}^+ + Z_{d2}^+$, $Z_2^- = Z_{s2}^- + Z_{d2}^-$ and $Z_2^0 = Z_{s2}^0 + Z_{d2}^0$

$$\left.\begin{array}{l} I_f^+ = -\dfrac{V_{fp}^+}{(Z_1^+) \| (Z_2^+)} \\ I_f^- = -\dfrac{V_{fp}^-}{(Z_1^-) \| (Z_2^-)} \\ I_f^0 = -\dfrac{V_{fp}^0}{3Z_f + (Z_1^0) \| (Z_2^0)} \end{array}\right\} (5)$$

$$I_2^+ = \frac{I_f^+ Z_1^+}{(Z_1^+)+(Z_2^+)}$$
$$I_2^- = \frac{I_f^- Z_1^-}{(Z_1^-)+(Z_2^-)} \quad (6)$$
$$I_2^0 = \frac{I_f^0 Z_1^0}{(Z_1^0)+(Z_2^0)}$$

$$I_1^+ = \frac{I_f^+ Z_2^+}{(Z_1^+)+(Z_2^+)}$$
$$I_1^- = \frac{I_f^- Z_2^-}{(Z_1^-)+(Z_2^-)} \quad (7)$$
$$I_1^0 = \frac{I_f^0 Z_2^0}{(Z_1^0)+(Z_2^0)}$$

Now the sequence components of the current contribution from the two buses are expressed in (6) and (7) respectively. Accordingly, the sequence components of the bus voltages become

$$V_1^+ = -I_1^+ Z_{s1}^+ = -\frac{I_f^+ Z_2^+ Z_{s1}^+}{(Z_1^+)+(Z_2^+)}$$
$$V_1^- = -I_1^- Z_{s1}^- = -\frac{I_f^- Z_2^- Z_{s1}^-}{(Z_1^-)+(Z_2^-)} \quad (8)$$
$$V_1^0 = -I_1^0 Z_{s1}^0 = -\frac{I_f^0 Z_2^0 Z_{s1}^0}{(Z_1^0)+(Z_2^0)}$$

$$V_2^+ = -I_2^+ Z_{s2}^+ = -\frac{I_f^+ Z_1^+ Z_{s2}^+}{(Z_1^+)+(Z_2^+)}$$
$$V_2^- = -I_2^- Z_{s2}^- = -\frac{I_f^- Z_1^- Z_{s2}^-}{(Z_1^-)+(Z_2^-)} \quad (9)$$
$$V_2^0 = -I_2^0 Z_{s2}^0 = -\frac{I_f^0 Z_1^0 Z_{s2}^0}{(Z_1^0)+(Z_2^0)}$$

*2. Symmetrical fault condition*

The analysis of symmetrical fault condition is different from the unsymmetrical fault condition. The decomposition of phase voltage and current under symmetrical fault condition is not applicable. Under such fault condition the balanced and symmetrical relationship between phase voltages and currents are not violated. Thus only positive sequence network is helpful for such fault analysis. Here, under such fault condition fault current $I_f^+$ is evaluated as (10).

$$I_f^+ = -\frac{V_{fp}^+}{Z_f + (Z_{s1}^+ + Z_{d1}^+) \parallel (Z_{s2}^+ + Z_{d2}^+)} = -\frac{V_{fp}^+}{Z_f + Z_1^+ \parallel Z_2^+} \quad (10)$$

Now the respective positive sequence current contributions from the two buses are

$$I_1^+ = \frac{I_f^+ Z_2^+}{(Z_1^+)+(Z_2^+)} \quad (11) \qquad I_2^+ = \frac{I_f^+ Z_1^+}{(Z_1^+)+(Z_2^+)} \quad (12)$$

Accordingly, positive sequence bus voltages become

$$V_1^+ = -I_1^+ Z_{s1}^+$$
$$= -\frac{I_f^+ Z_2^+ Z_{s1}^+}{(Z_1^+)+(Z_2^+)} \quad (13)$$

$$V_2^+ = -I_2^+ Z_{s2}^+$$
$$= -\frac{I_f^+ Z_1^+ Z_{s2}^+}{(Z_1^+)+(Z_2^+)} \quad (14)$$

*b) External fault*

In this section the fault occurrence is considered outside the suggested MG as depicted in Fig.2. Here, in case of unsymmetrical fault occurrence all the symmetrical components are required for fault analysis. Whereas, in symmetrical fault condition, only positive sequence component is considered. According to Fig. 2 the positive, negative and zero sequence voltage components of the buses are $V_1^+$, $V_1^-$, $V_1^0$ and $V_2^+$, $V_2^-$, $V_2^0$. Rearranging the equation the symmetrical components of the fault current between bus 1 and 2 are evaluated as (15)

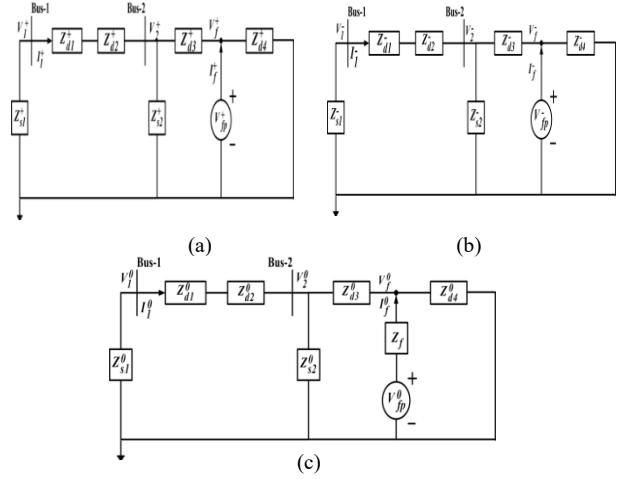

Fig.2 (a) Positive sequence network for external fault condition, (b) negative sequence network for external fault condition and (c) zero sequence network for external fault condition.

$$V_1^+ - V_2^+ = I_1^+ (Z_{d1}^+ + Z_{d2}^+)$$
$$V_1^- - V_2^- = I_1^- (Z_{d1}^- + Z_{d2}^-) \quad (15)$$
$$V_1^0 - V_2^0 = I_1^0 (Z_{d1}^0 + Z_{d2}^0)$$

## III. FAULT TYPE IDENTIFICATION AND ISOLATION

In this section the proposed fault identification and fault isolation topology is illustrated in detail. Generally, in any power system protection design it is very important to differentiate the load switching and fault condition in order to avoid false tripping. Thus, here the locally measured phase voltages and currents are considered at first to differentiate between no fault and with fault condition. The measured voltage $V_{actual}$ and current $I_{actual}$ are compared with their respective rated values to compute the mismatch.

$$V_{er} = V_{actual} - V_{rated}$$
$$I_{er} = I_{actual} - I_{rated} \quad (16) \qquad V_{er}\% \leq V_{threshold}$$
$$I_{er}\% \geq I_{threshold} \quad (17)$$

According to (16) the computed voltage and current errors are $V_{er}$ and $I_{er}$ respectively. $V_{rated}$ and $I_{rated}$ represents the rated value of voltage and current. The percentage of this mismatch is further compared with some pre considered thresholds to understand the no fault and with fault condition. Whenever the voltage mismatch percentage becomes lesser and current mismatch percentage become higher than the pre considered threshold (17) fault is located. Further the identification of the exact type of fault is carried out using symmetrical component analysis as presented in the following flow in Fig.3.

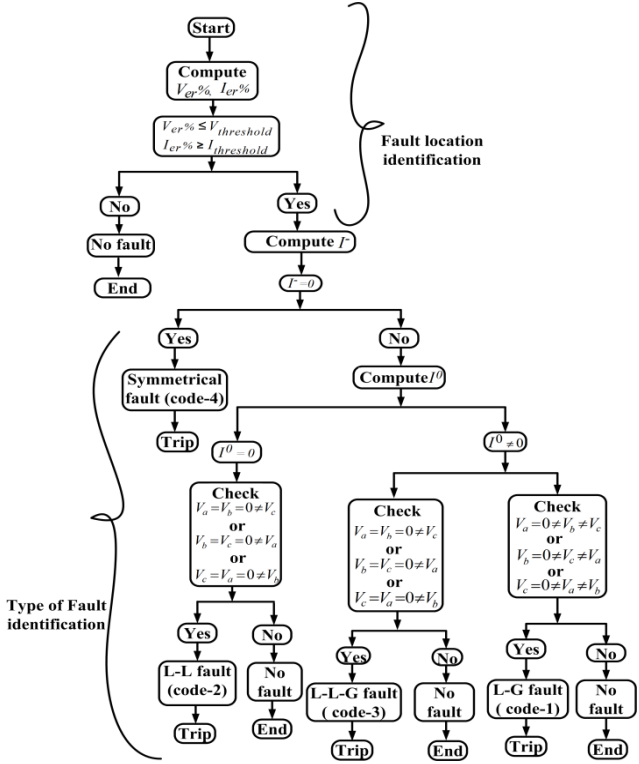

Fig.3 Flow chart of fault locate, isolation and exact fault type identification.

Majority of the available protection schemes are directly or indirectly dependent upon the commutation channel to locate and identify faults. These communication based protection topologies often suffer from delay or failure of information circulation which may result in disaster in case of severe faults. In order to address such concern the proposed topology is implemented without including any communication between the different sections of power system model.

In practical application isolation of the fault affected zone from existing interconnected system is required to restrict the fast propagation of fault currents. Thus after locating and identifying the exact fault immediate tripping signal is initiated by the solid state relay. The tripping signal communication to the static switch exhibits the isolation operation to restrict the propagation of fault current.

IV. MICROGRID TESTBED

The proposed protection topology is tested on a low-voltage distribution level mesh configured MG, modeled using MATLAB/Simulink platform as depicted in Fig. 4. The MG operating voltage is 415V at 50 Hz. The MG consist of total 5 buses, transformer, loads of different type (including ac and dc), 3 DERs (PV array, fuel cell and battery storage), DC to AC converters and LC filters. The utility grid is connected to the MG through a static switch in order to achieve grid connected and islanded mode operation. All distribution line length is considered as 1 km. Here at the beginning and end of each distribution line the local measurement units and solid state relay driven static switch are incorporated. The specific rating about system components are furnished in Table-I.

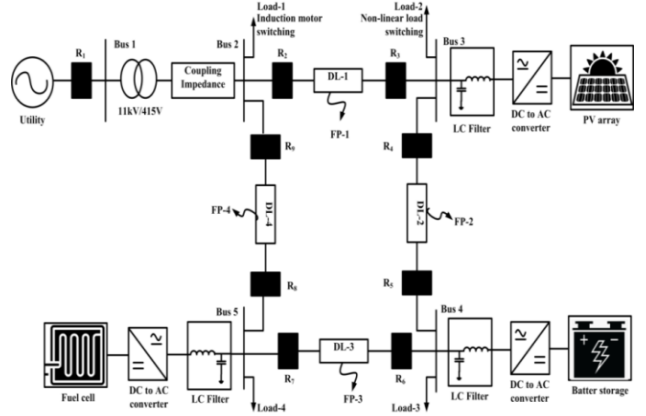

Fig.4 Considered microgrid test bed.

| TABLE-I Specification of system components | | |
|---|---|---|
| System component | | Specification |
| Utility grid | | 500kW, 11kV, 50Hz |
| Transformer | | 1 MVA, 11kV/415V |
| DER | PV array | 63.18kW, 45 V |
| | Fuel cell | 10 kW, 45 V |
| | Battery storage | 50 Ah, 357 V |
| R-L load | Load-1, Load-2 | 20 kW, 0.85 pf |
| | Load-3 | 12 kW, 0.8 pf |
| | Load-4 | 5 kW, 0.9 pf |
| Induction motor | | 415 V, 5 KW |
| Non-linear load (Resistive load) | | 1 kW |

V. SIMULATION RESULTS AND DISCUSSION

The performance of the proposed protection scheme is investigated under different power system scenarios in the MG test bed as depicted in Fig.4. The detailed performance analysis with simulation test results is presented in the following section.

a. *Fault testing in grid connected situation*

The fault locations are chosen as FP-1, FP-2, FP-3 and FP-4 which are at appearing at the mid-point of the distribution lines DL-1, DL-2, DL-3, DL-4 respectively. The type of faults with their instant of occurrence is included in Table-II. Here the different type of faults is identified in term of different numerical values (like for L-G fault identification code is '1'). After fault detection and identification DL-1, DL-2, DL-3, DL-4 associated respective solid state relays initiates a tripping signal for fast isolation of the fault affected distribution lines.

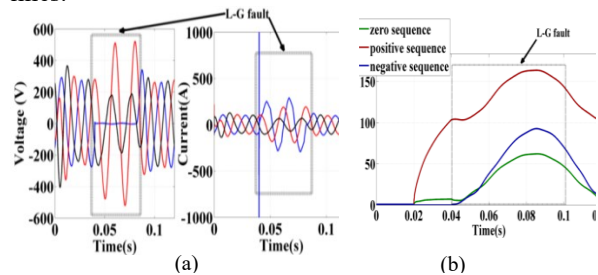

(a)       (b)

Fig.5 (a) Overall voltage and current waveform of DL-1 and (b) current decomposition waveform of DL-1.

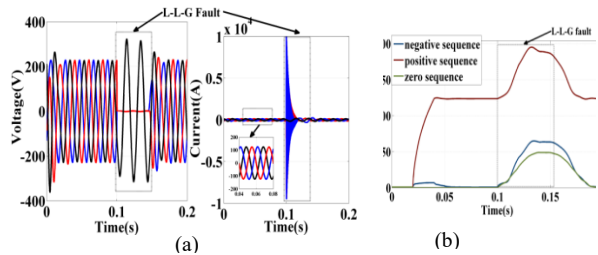

(a)       (b)

Fig.6 (a) Overall voltage and current waveform of DL-2 and (b) current decomposition waveform of DL-2.

To represent the fault studies, a L-G fault FP-1 is externally created at 0.04 s (as depicted in Fig.4) at the midpoint of DL-1 without having any fault resistance ($R_f$ =0). The corresponding current and voltage waveforms during such fault situation are depicted in Fig.5 (a).The decomposed positive, negative and zero sequence current components are presented in Fig.5 (b) as extracted from the local measurement. These nature of current and voltage waveforms are useful for type of fault identification. Similarly, Fig.6 (a) depicts the situation of voltage and current under FP-2 (L-L-G) fault at the midpoint of DL-2. Here also positive, negative and zero sequence current components are present during fault condition as presented in Fig.6 (b).For both FP-1 and FP-2 condition as ground is available thus apart from sudden rise in current and abnormal fall in voltage sequence currents are present only during faults. Unlike this for FP-3 (L-L) as ground is not available, only positive and negative sequence current decomposition takes place during fault. Fig.7 (a) clearly depicts the voltage and current wave shapes of DL-3. Accordingly, Fig.7 (b) signifies the difference in decomposition of fault current between FP-2 and FP-3. Lastly the major fault i.e. FP-4 (L-L-L-G) is carried out at the mid-point of DL-4. Since it is symmetrical fault only massive hike in positive sequence component and is notable (Fig.8 (b)).

In addition to this it is clearly visible that when FP-1, FP-2, FP-3 and FP-4 are tested one by one at DL-1, DL-2, DL-3 and DL-4. Here the selective isolation of respective relay has restricted the fault current proliferation. As a result except the fault affected bus other neighboring buses remain unaffected.

However the negative and zero sequence components remain unavailable.

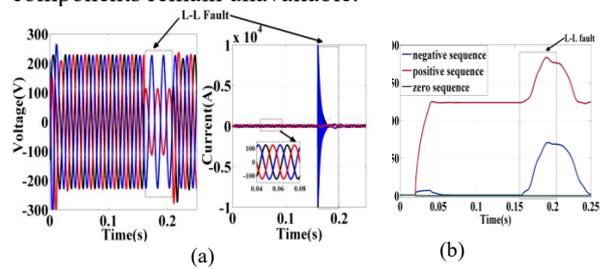

(a)       (b)

Fig.7 (a) Overall voltage and current waveform of DL-3 and (b) current decomposition waveform of DL-3.

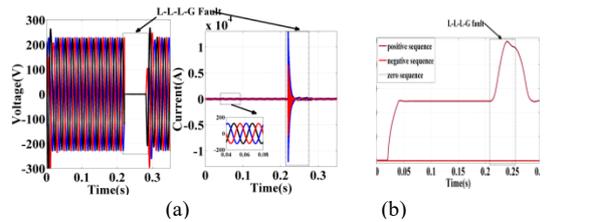

(a)       (b)

Fig.8 (a) Overall voltage and current waveform of DL-4 and (b) current decomposition waveform of DL-4.

| TABLE-II Tested fault details | | | |
|---|---|---|---|
| *Fault point* | *Instant of fault* | *Code* | *Relay* |
| FP-1 at midpoint DL-1 (L-G) | 0.04 s | 1 | $R_2, R_3$ |
| FP-3 at midpoint DL-3 (L-L) | 0.1 s | 2 | $R_6, R_7$ |
| FP-2 at midpoint DL-2 (L-L-G) | 0.16 s | 3 | $R_4, R_5$ |
| FP-3 at midpoint DL-4 (L-L-L-G) | 0.21 s | 4 | $R_8, R_9$ |

b. *Fault testing in islanded situation with fault resistance*

The fault resistance is an important parameter for locating and identifying the fault as fault current magnitude decreases sharply due to inclusion of fault resistance. Thus, here a range of fault resistances between 1Ω to 20Ω is considered in islanded mode, to examine the ability to locate, isolate and identify the faults. It has been observed that even for the worst case condition i.e. different faults with the fault resistance of 20Ω fault location, identification and isolation has been achieved properly. In order to determine the threshold values of voltage and current the two extreme situations i.e. L-G fault with 20Ω fault resistance and symmetrical fault with 20Ω in islanded mode operation are considered .The response time of relay for identification and isolation using the proposed topology for different fault cases both in grid connected and islanded mode operation is presented in Table-III.

c. *Sensitivity testing with no fault cases*

Any protection scheme should be capable to differentiate the no fault and fault situations. In this regard, the proposed protection scheme is further tested with a few sensitive load switching conditions to examine the performance of the proposed topology.

The switching time as well as rating of various loads switched at DL-1 is furnished in Table-IV. It can be observed that the maximum delay in relay response i.e. 0.0025 s is obtained for L-L-L-G fault situation in islanded mode operation when 20 Ω fault resistance. Thus it can be stated based on Fig.9 that while such type of sensitive switching are carried out the proposed topology does not execute false tripping.

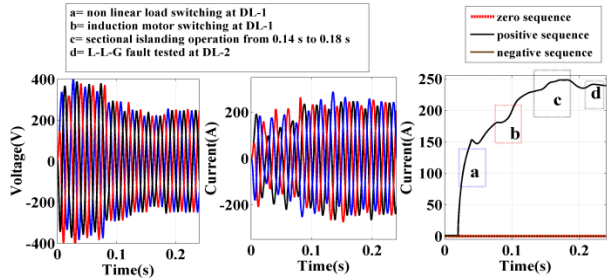

Fig.9 Overall voltage current and symmetrical current components of DL-1 with no fault switching cases.

### TABLE-III
### Response time of the proposed protection scheme under different faults with different fault resistance

| Type of fault | Location | Response time in grid connected mode | Response time in islanded mode | $R_f$ | Code |
|---|---|---|---|---|---|
| L-G | Mid point of DL-1 | 0.0001 s | 0.0008 s | 20 Ω | 1 |
| L-L-L-G | Mid point of DL-4 | 0.0003 s | **0.0025** s | 20 Ω | 4 |
| L-L | Mid point of DL-3 | 0.0006 s | 0.0001 s | 10 Ω | 2 |
| L-L-G | Mid point of DL-2 | 0.0002 s | 0.0001 s | 12 Ω | 3 |
| L-G | Mid point of DL-1 | 0.000005 s | 0.000003 s | 8 Ω | 1 |
| L-L-G | Mid point of DL-2 | 0.00004 s | 0.00002 s | 4 Ω | 3 |
| L-L-L-G | Mid point of DL-4 | 0.0001 s | 0.00006 s | 3 Ω | 4 |

### TABLE-IV
### Sensitivity Testing With No-Fault Cases

| Mode of operation | Instant of switching | Type of switching | Rating | Fault identification |
|---|---|---|---|---|
| Islanded | 0.04 s | non-linear load | 2 kw | No fault |
| Islanded | 0.08 s | induction motor | 5 kw,0.85 pf | No fault |
| Grid connected | (0.14-0.18) s | sectional islanding | - | No fault |
| Islanded | 0.2 s | L-L-G at DL-2 | - | No fault (at DL-1) |

## VI. CONCLUSION

A simple and faster fault locate, identify and isolation scheme is proposed in this paper based on local measurements. Based on the simulation results it can be concluded that

1. The proposed protection scheme is compatible both in grid connected and islanded MG condition in order to fault locate, identify and isolation.
2. The proposed scheme is designed only based on local measurements. In order to avoid the phenomena of false tripping threshold of voltage and current is determined after examining every critical operating condition.
3. Additionally, comparatively faster fault isolation is achieved as solid state relays are incorporated instead of over-current or differential relays and traditional circuit breakers are substituted by static switches.

The extensive test results establish that the accuracy and fast responding time of the proposed scheme is capable to provide a reliable protection measure in low voltage ac MG.